# Si-induced superconductivity and structural transformations in DyRh$_4$B$_4$

A. Köhler[1], G. Behr[1], G. Fuchs[1#], K. Nenkov[1], L.C. Gupta[2*]

[1] IFW Dresden, Leibniz-Institut für Festkörper- und Werkstoffforschung Dresden e.V.,
 P.O. Box 270116, D-01171 Dresden, Germany
[2] Max Planck Institute for the Physics of Complex Systems, D-01187, Dresden, Germany

**Abstract**

DyRh$_4$B$_4$ has been known to crystallize in the primitive tetragonal (*pt*)-structure and to exhibit a ferromagnetic transition at 12 K, the highest magnetic transition temperature in the entire series of the *R*Rh$_4$B$_4$ materials [1]. We show here that our silicon-added samples of the nominal composition DyRh$_4$B$_4$Si$_{0.2}$ exhibit superconductivity below T$_c$ ~ 4.5 K and an antiferromagnetic transition below T$_N$ ~ 2.7 K. The 12 K transition observed in the *pt*-DyRh$_4$B$_4$ is completely suppressed. Our annealed samples mainly consist of domains of the chemical composition DyRh$_{3.9}$B$_{4.2}$Si$_{0.08}$. These domains contain two crystallographic phases belonging to the body-centred tetragonal (*bct*)-structure and the orthorhombic (*o*)-structure. We have reasons to suggest that superconductivity and antiferromagnetic ordering arise from *bct*- DyRh$_4$B$_4$ phase and, therefore, coexist below T$_N$ ~ 2.7 K.

\# Corresponding author:
    G. Fuchs
    Leibniz-Institut für Festkörper- und Werkstoffforschung Dresden
    P.O. Box 270116, D-01171 Dresden, Germany
    Fax: +49-351-4659-490
    Tel.: +49-351-4659-538
    E-mail: fuchs@ifw-dresden.de

[*]Visiting scientist

## 1. Introduction

Kuz'ma et al [2] reported the formation of the borides $R$M$_4$B$_4$, with $R$ = rare earths, Y and Th and $M$ = Co, in the CeCo$_4$B$_4$-type primitive tetragonal (*pt*) structure, shown in Fig. 1a. These materials exhibit neither a superconducting nor a magnetic transition above 1.5 K. Matthias et al. [3] reported the Rh-based series of ternary materials $R$Rh$_4$B$_4$ crystallizing in the *pt*-CeCo$_4$B$_4$-structure. These materials generated great interest not only because the two nonmagnetic members of this series, LuRh$_4$B$_4$ and YRh$_4$B$_4$, superconduct with high T$_c$ (> 10 K) but also because several members of this series (with $R$ = Nd, Sm, Er, Tm) exhibit simultaneously both superconductivity and long range magnetism. These materials, therefore, have been studied extensively for the interplay of superconductivity and magnetism [1]. Materials with composition $R$M$_4$B$_4$ crystallizing in structures *other than the pt*-CeCo$_4$B$_4$-type have also been reported. For example, $R$Ru$_4$B$_4$ crystallize in the *bct*-tetragonal structure, see Fig. 1b; none of them, except LuRu$_4$B$_4$ (T$_c$ ~ 1.8 K), exhibits superconductivity [4]. "Mixed materials" with composition $R$(Rh$_{1-x}$Ru$_x$)$_4$B$_4$ (x ~0.15) crystallize, essentially as single phase, in the *bct*-LuRu$_4$B$_4$ structure; those with heavy rare earths ($R$ = Lu, Tm, Er, Ho, Dy) superconduct [4].

*bct*-ErRh$_4$B$_4$ has been synthesized and studied. It does exhibit superconductivity (T$_c$ ~ 7.8 K) as in *pt*-ErRh$_4$B$_4$. However, instead of ferromagnetic ordering in *pt*-ErRh$_4$B$_4$, *bct*-ErRh$_4$B$_4$ undergoes an antiferromagnetic ordering below T$_N$ = 0.65 K [5]. *bct*-ErRh$_4$B$_4$ has also been produced by carbon-doping (T$_c$ ~ 7.5 K and T$_N$ ~ 0.9 K) [6]. LuRh$_4$B$_4$ was reported forming in yet another structure, an orthorhombic structure known in literature as *o*-LuRh$_4$B$_4$ [7]. The structure is shown in Fig. 1c. Some materials of the composition $R$Rh$_4$B$_4$ ($R$ = heavy rare earths such as Ho, Er, Tm, Yb, Lu) have been synthesized crystallizing in the *o*-LuRh$_4$B$_4$ and their magnetic and superconducting properties have been reported [8]. DyRh$_4$B$_4$ and YRh$_4$B$_4$ have been produced in the *bct*-LuRu$_4$B$_4$ structure using high pressure synthesis route [9]. A significant result of this work was the observation of superconductivity in *bct*-DyRh$_4$B$_4$ at T$_c$

~ 4.5 K. A ferromagnetic transition at a rather high temperature, $T_M$ ~ 40 K, was also observed in *bct*-DyRh$_4$B$_4$ [9]. The superconducting and magnetic ordering temperatures reported for ErRh$_4$B$_4$ and DyRh$_4$B$_4$ are listed in Table 1.

Recently, superconductivity was reported at $T_c$ ~ 10 K in multiphase materials of the quaternary system Y-Rh-B-Si [10]. As we mentioned above, *bct*-DyRh$_4$B$_4$ and *bct*-ErRh$_4$B$_4$ have superconducting and magnetic properties that are remarkably different from those of *pt*-DyRh$_4$B$_4$ and *pt*-ErRh$_4$B$_4$. Such considerations motivated us to try another chemical route, namely, the effect of introducing Si, to promote the structural transformation $pt \rightarrow bct$ among the members of the series RRh$_4$B$_4$. DyRh$_4$B$_4$ was chosen in the present work for such studies.

## 2. Experimental

Polycrystalline samples of nominal composition DyRh$_4$B$_4$Si$_x$ (x = 0, 0.1 and 0.2) were synthesized by arc-melting. The as-prepared ingots were wrapped in tantalum foils and annealed at 1100 °C for 72 hours in an evacuated silica tube. Final composition and the phases present in the samples were checked and studied by electron probe microanalysis. Powder X-ray diffraction was used to determine the crystal structures of the major phases present in our materials. Magnetic and superconducting properties of the as-cast and the annealed samples were measured using Quantum Design PPMS and SQUID magnetometers.

## 3. Results and discussion

The AC-susceptibility measurements, shown in Fig. 2, of the sample DyRh$_4$B$_4$ (with no added silicon) reveal the well known ferromagnetic transition at ~ 12 K in this material [3]. Inset in Fig. 2 shows the corresponding hysteresis curve in the ferromagnetic state at T = 2 K. The temperature dependence of the inverse DC susceptibility of this samples shows Curie-Weiss behavior between 300 K and 50 K with a Curie-Weiss temperature of 18.0 K and an effective magnetic moment of $\mu_{eff} = 10.4$ $\mu_B$ per $Dy^{3+}$ ion which is slightly below the free ion value of 10.63 $\mu_B$ per Dy ion. Similar data have been reported already for pt-DyRh$_4$B$_4$ [3].

In Fig. 2, there are, besides the main peak at 12 K, two other maxima at 23 K and 37 K in the AC-susceptibility curve. We shall comment upon them below when we discuss the results of the microanalysis of the samples.

AC-susceptibility measurements of the as-cast DyRh$_4$B$_4$Si$_{0.2}$ sample, Fig. 3a, show a superconducting transition at T = 4.5 K with a large diamagnetic response. As can be seen from Fig. 3a, the ferromagnetic transition (at ~12.3 K) observed in silicon-free sample has vanished. Further, we see a peak at 37 K in this silicon-added as-cast sample also[1]. It is remarkable that this magnetic transition vanishes in the annealed sample, see Fig. 3b. The volume fraction of the superconducting part in this sample was determined as ~ 50% by means of the slope of the field dependent magnetization curve in the Meißner state.

The sample DyRh$_4$B$_4$Si$_{0.1}$ with lower silicon content also superconducts below 4 K. However the superconducting volume fraction in this case is only 7%. A magnetic transition is observed at $T_M = 12$ K in the as-cast sample indicating that a small fraction of pt-DyRh$_4$B$_4$ phase still continues to exist in the material. This transition disappears in the annealed sample and the magnetic transition observed below $T_N = 2.7$ K survives. Thus, the two annealed samples DyRh$_4$B$_4$Si$_{0.1}$ and DyRh$_4$B$_4$Si$_{0.2}$ have similar superconducting and magnetic properties, except for the low superconducting volume fraction in the sample DyRh$_4$B$_4$Si$_{0.1}$. In the following we shall focus our attention on the material DyRh$_4$B$_4$Si$_{0.2}$ only.

Electron probe microanalysis of the as-cast and the annealed samples has been carried out. The insets in Figs. 3a and 3b show the backscattered electron (BSE) images of the superconducting DyRh$_4$B$_4$Si$_{0.2}$ samples (as-cast sample in Fig. 3a and the annealed sample in Fig. 3b). The gray regions, which form the most part of the samples, have the composition

---

[1] As pointed out earlier, a similar peak is observed in our as-cast silicon free sample DyRh$_4$B$_4$, Fig. 2.

DyRh$_{3.9}$B$_{4.6}$Si$_{0.08}$ in the as-cast and DyRh$_{3.9}$B$_{4.2}$Si$_{0.08}$ in the annealed sample. Added silicon is not fully incorporated (the assimilated silicon is lower than the silicon taken to start with). White regions correspond to the compositions DyRh$_{3.1}$B$_{2.5}$Si$_{0.05}$ in the as-cast and DyRh$_{3.0}$B$_{2.0}$Si$_{0.35}$ in the annealed samples. This secondary phase can be identified as a related phase to the mother compound DyRh$_3$B$_2$ with monoclinic ErIr$_3$B$_2$-type structure [11] (also see Table 2.9 in [8]). During the annealing process, amount of this 132-phase is considerably reduced which is obvious by comparing the two BSE images (insets of Figs. 3a and 3b). Simultaneously the magnetic transition at T$_M$ = 37 K (Fig. 3a) vanishes. Thus we conclude that this magnetic transition is caused by the secondary phase inclusions. This thesis is supported by our magnetic measurements on samples of nominal composition of DyRh$_{3.0}$B$_{2.0}$Si$_{0.2}$. Similar behaviour has been observed in ErIr$_4$B$_4$ compound [11]. Likewise, the additional maxima in Fig. 2 are attributed to the presence of some other minority impurity phases.

The magnetic phase diagram of DyRh$_4$B$_4$Si$_{0.2}$ in the *H-T* plane shown in Fig. 4 reveals bulk superconductivity in the measured temperature range between $T_c$ and 2 K below the upper critical field $H_{c2}(T)$. The magnetization in the superconducting state was investigated as function of temperature in the zero-field cooling (*zfc*) and in the field cooling (*fc*) mode applying magnetic fields up to 0.1 T. A peak in the *fc* branch of the *dc* susceptibility appearing at nearly the same temperature *T\** for each applied field announce magnetic ordering at this temperature, i.e. $T_N = T^* \approx 2.7$ K (see Fig. 4). The irreversibility field determined from these data was found to increase with decreasing temperature in the paramagnetic state. The range between $H_{irr}$ and $H_{c2}$ in which the magnetization remains reversible occupies a large range of the magnetic phase diagram indicating the weak flux pinning properties of the investigated sample.

According to Fig. 4, superconductivity and antiferromagnetism are found to coexist in DyRh$_4$B$_4$Si$_{0.2}$ at temperatures below $T_N \approx 2.7$ K. Note that the slope d$H_{c2}$/d$T$ of the $H_{c2}(T)$ curve becomes larger below $T_N$, i.e. the superconducting properties are observed to improve in the antiferromagnetically ordered state. The coexistence of superconductivity and antiferromagnetism was additionally confirmed by magnetization vs. field measurements performed at temperatures below $T_N$. The magnetization data plotted in Fig. 5 show both superconductivity at magnetic fields below about 0.4 T (including the Meissner effect at very low fields and a hysteretic behaviour due to flux pinning in the field range up to about 0.2 T) and the non-hysteretic antiferromagnetic "background" which becomes visible at fields $H > H_{c2}$. The highly non-linear $M(H)$ dependence above $\mu_o H \approx 1$T (see inset in Fig 5) corresponds to the Brillouin function describing the magnetization of the paramagnetic phase at a large ratio $\mu_B H/kT$.

X-ray powder diffraction pattern of the annealed $DyRh_4B_4Si_{0.2}$ sample is shown in Fig. 6a. This pattern consists of the expected pattern of the *bct*-$LuRu_4B_4$ crystal structure type (blue reference diffraction pattern shown in Fig. 6b) and it also has intense additional diffraction lines, for example, at $2\theta = 36.5°$ and $2\theta = 39°$. We identified these as originating from the orthorhombic polytype of $DyRh_4B_4$ (red reference pattern of *o*-$LuRh_4B_4$ in Fig. 6b, the appropriate crystal structure is drawn in Fig. 1c). It is to be pointed out here that the occurrence of the *o*-$LuRh_4B_4$ type structure in $DyRh_4B_4$ has not been reported so far in literature. Since we find a mixture of the two polytypes *bct*-$DyRh_4B_4$ and *o*-$DyRh_4B_4$, closely related to each other, we conclude that the assimilation of silicon in $DyRh_4B_4$ causes the formation of both of these polytypes. This is in contrast to the effect of C added to $ErRh_4B_4$ wherein only the *bct*-type structure is stabilized [6]. It appears to us that silicon may be entering in *more than one* crystallographic sites in $DyRh_4B_4$. In dependence on which crystallographic site silicon is incorporated, the *bct*-phase or the *o*-phase will be favoured. This generates in the sample a mixture of the two phases. It would be of interest to study further this phenomenon, namely, *an impurity atom stabilizing two crystallographic structures* in the $R$Rh$_4$B$_4$ system.

We attribute the superconducting transition at 4.4 K to the *bct*-DyRh$_4$B$_4$ for the following reasons: It is known [4] that Dy(Rh$_{0.85}$Ru$_{0.15}$)$_4$B$_4$ crystallizes in the *bct*-LuRu$_4$B$_4$ structure and superconducts at T$_c$ ~ 4.0 K. Further, according to the recent work [9], *bct*-DyRh$_4$B$_4$ (high pressure route) exhibits superconductivity at ~ 4.5 K. In both cases, T$_c$ is close to that found in our work. Similar results are available in the case of ErRh$_4$B$_4$. *bct*-ErRh$_4$B$_4$ has been obtained through two routes, high pressure synthesis [5] and incorporation of C [6]. Superconducting and magnetic properties of the resulting *bct*-ErRh$_4$B$_4$ are nearly same. These two examples suggest that the observed superconducting and magnetic properties are specific to the phase rather than the route followed to realize that phase and this is what it should be. Thus it is quite reasonable to propose that it is the *bct*-phase of DyRh$_4$B$_4$ that is responsible for the observed superconductivity in DyRh$_4$B$_4$Si$_{0.2}$. Further support to our proposal comes from our own studies on the annealed material DyRh$_4$B$_4$Si$_{0.1}$. X-ray powder diffraction of this sample shows that it contains a *higher* fraction of the orthorhombic phase while the superconducting volume fraction is much lower than that in the DyRh$_4$B$_4$Si$_{0.2}$ sample. Magnetic response of DyRh$_4$B$_4$Si$_{0.1}$ is also much weaker than that of DyRh$_4$B$_4$Si$_{0.2}$. This observation motivates us to suggest that the magnetic transition that we see in DyRh$_4$B$_4$Si$_{0.2}$ also arises from the *bct*-DyRh$_4$B$_4$, implying thereby coexistence of superconductivity and antiferromagnetism in *bct*-DyRh$_4$B$_4$ below T$_N$ ~ 2.7 K. In Table 2, our data are compared with reported data for *pt* and *bct* DyRh$_4$B$_4$ [1,4,9]. More work is required to be able to comment upon the superconducting/magnetic properties of *o*-DyRh$_4$B$_4$.

## 4. Conclusions

In conclusion, we report here the results of our studies of incorporation of Si in DyRh$_4$B$_4$. Our work shows that samples of nominal composition DyRh$_4$B$_4$Si$_{0.2}$ have two structural phases, $bct$-DyRh$_4$B$_4$ and $o$-DyRh$_4$B$_4$. We attribute the observed superconductivity at T$_c$ ~ 4.4 K and antiferromagnetic transition at T$_N$ ~ 2.7 K in DyRh$_4$B$_4$Si$_{0.2}$ to the $bct$-DyRh$_4$B$_4$ which implies coexistence of superconductivity and magnetic order below 2.7 K in this phase.

As we do not see any magnetic transition in the vicinity of 37 K in our annealed samples which have $bct$-DyRh$_4$B$_4$, it appears to us that the reported magnetic transition at 40 K in the $bct$-DyRh$_4$B$_4$ [9] is most probably due to the presence of the DyRh$_3$B$_2$ which is known to exhibit [11] a magnetic transition at ~ 37 K.

It would be of interest to extend this study to other members of the $R$Rh$_4$B$_4$-series, namely, to try to incorporate suitable elements in the matrix $R$Rh$_4$B$_4$ and study the structural transformation(s) and superconducting and magnetic properties of the resulting phases of $R$Rh$_4$B$_4$.

**Acknowledgements**

The authors would like to thank S. Müller-Litvanyi, S. Pichl and A. Ostwaldt for technical support and J. Dshemuchadse, T. Leisegang and W. Löser for fruitful discussions. The financial support of the Deutsche Forschungsgemeinschaft within the SFB 463 „Rare-Earth Intermetallics: Structure, Magnetism and Transport" is grateful acknowledged. LCG thanks MPIPKS for providing support and hospitality during his several visits. He also thanks Humboldt Foundation for the support during some of his visits to MPIPKS.

**Tables**

Table 1
Superconducting ($T_c$) and magnetic transition temperatures ($T_N$, $T_M$) of ErRh$_4$B$_4$ and DyRh$_4$B$_4$

|  | **ErRh$_4$B$_4$** | | | **DyRh$_4$B$_4$** | | |
|---|---|---|---|---|---|---|
|  | *pt* | *bct* | *o* | *pt* | **bct* | *♦bct* |
| **T$_c$ / K** | 8.7 | 7.8 | 4.5 |  | 4.0 | 4.5 |
| **T$_N$ / K** |  | 0.65 | 0.3 |  | 1.5 |  |
| **T$_M$ / K** | 0.9 |  |  | 12.0 |  | 40 |

$^*bct$-Dy(Rh$_{1-0.15}$Ru$_{0.15}$)$_4$B$_4$ [4], $^♦bct$-DyRh$_4$B$_4$ high pressure synthesis compound [9]; the orthorhombic crystal structure was reported in literature for ErRh$_4$B$_4$ but not for DyRh$_4$B$_4$.

Table 2
Superconducting ($T_c$) and magnetic transition temperatures ($T_N$, $T_M$) of DyRh$_4$B$_4$

| **DyRh$_4$B$_4$** | | | | | | |
|---|---|---|---|---|---|---|
|  | *pt* | *pt* | *ature*bct* | *♦bct* | *bct* | *o* |
| **T$_c$ / K** |  |  | 4.0 | 4.5 | **4.4** | < 2K |
| **T$_N$ / K** |  |  | 1.5 |  | **2.7** |  |
| **T$_M$ / K** | 12.0 | **12.3** |  | 40 |  |  |

$^*bct$-Dy(Rh$_{1-0.15}$Ru$_{0.15}$)$_4$B$_4$ [4], $^♦bct$-DyRh$_4$B$_4$ high pressure synthesis [9], our work in bold.

**Figure Captions**

**Figure 1**
Lattice polytypes of the materials $R$Rh$_4$B$_4$
a) CeCo$_4$B$_4$-type primitive tetragonal structure (45° rotated)
b) LuRu$_4$B$_4$-type body centered tetragonal structure
c) LuRh$_4$B$_4$-type orthorhombic structure
All polytypes consist of the same rare earth (orange balls) sub-lattice, but different orientations of the Rh (dark blue balls) and B (light blue balls) polyhedra, which is marked by different colours of the polyhedra; please note: the rare earth element is not inside the polyhedra but above in next layer; the unit cell is marked by a frame.

**Figure 2** (colour online)
AC-susceptibility of the DyRh$_4$B$_4$ as-cast sample,
inset: magnetic hysteresis curve at $T = 2$ K.

**Figure 3** (colour online)
AC-susceptibility of the a) as grown and b) annealed DyRh$_4$B$_4$Si$_{0.2}$ sample,
inset: backscattered electron (BSE) images.

**Figure 4** (colour online)
Temperature dependence of the upper critical field $H_{c2}$ of the annealed DyRh$_4$B$_4$Si$_{0.2}$ sample (solid line, blue circles) as determined from the onset of superconductivity of *ac* susceptibility data. Dotted line: antiferromagnetic ordering temperature $T_N$ corresponding to the temperature $T^*$ (red circles) at which a peak appears in the field cooling branch of the *dc* susceptibility; dashed line (open circles): irreversibility field $H_{irr}$. Coexistence of superconductivity and antiferromagnetism is observed below $T_N$.

**Figure 5** (colour online)
Field dependence of the magnetic moment of the annealed DyRh$_4$B$_4$Si$_{0.2}$ sample measured at $T = 2$K below $T_N = 2.7$ (see Fig. 4). Superconductivity in the magnetically ordered state is indicated by the diamagnetism at low applied fields (Meissner state) and the hysteretic character of the magnetization loop.

**Figure 6**
a) X-ray powder diffraction pattern (θ-2θ scan) of the annealed DyRh$_4$B$_4$Si$_{0.2}$
b) comparison of the peaks of the measured diffraction pattern (upper panel) with the peaks of the three reference patterns: *bct*-LuRu$_4$B$_4$ (blue), *o*-LuRh$_4$B$_4$ (red) and *pt*-YRh$_4$B$_4$ (green)

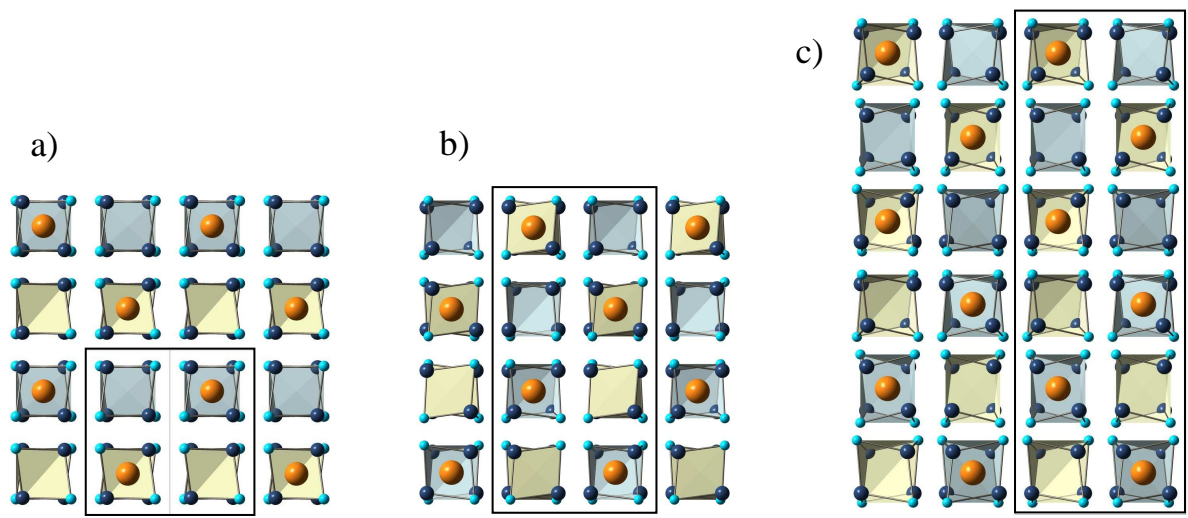

**Fig. 1**

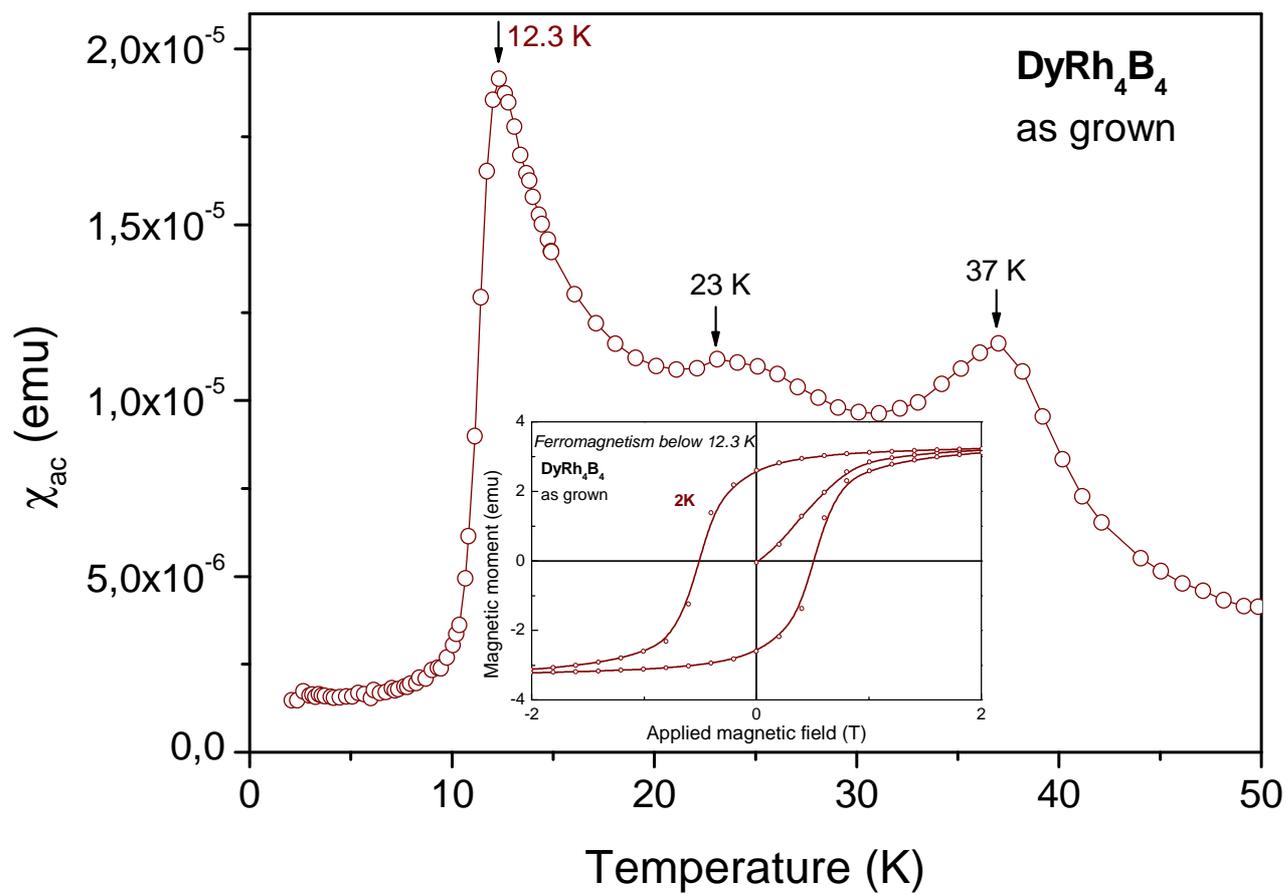

**Fig. 2**

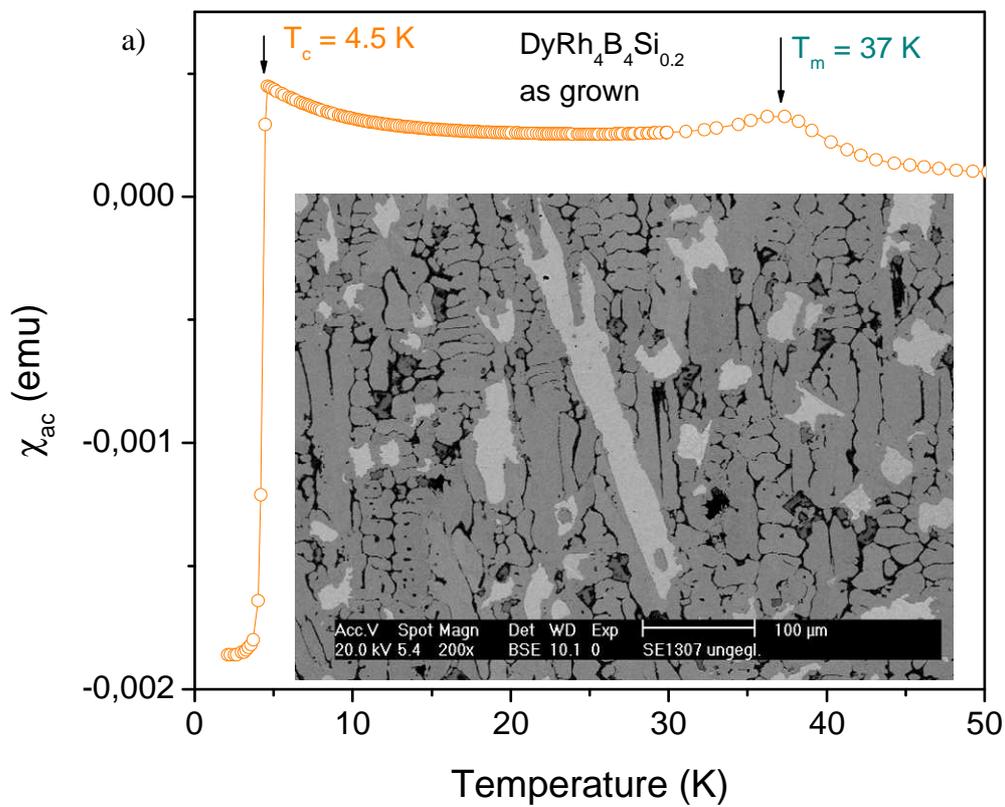

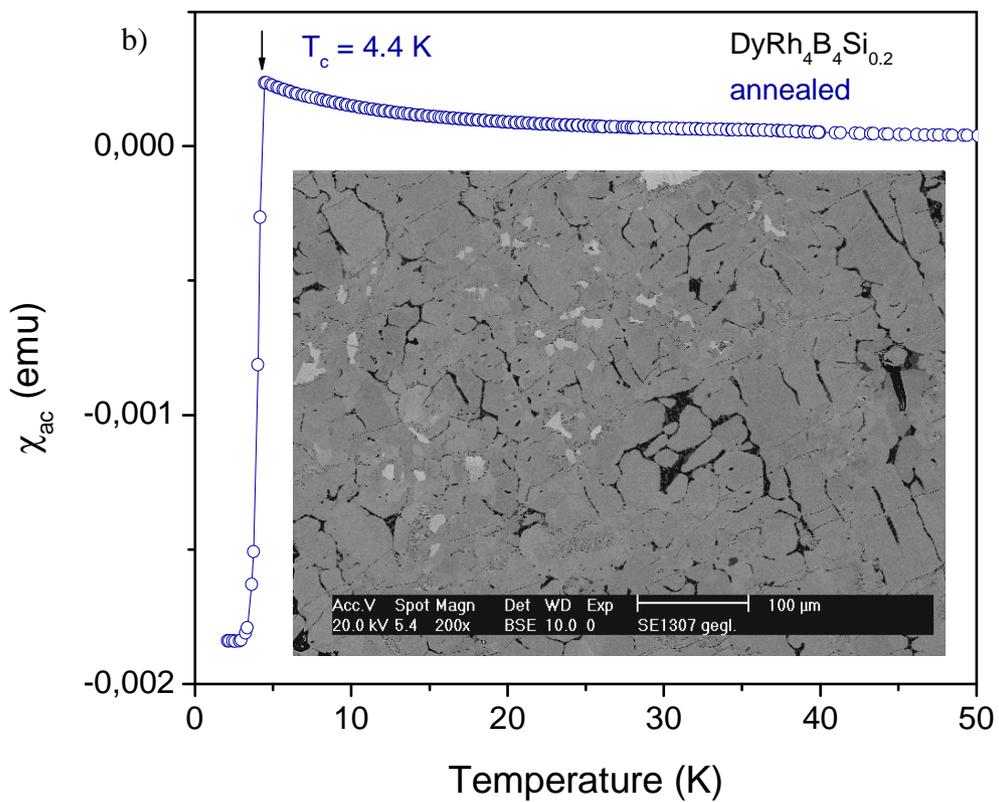

**Fig. 3**

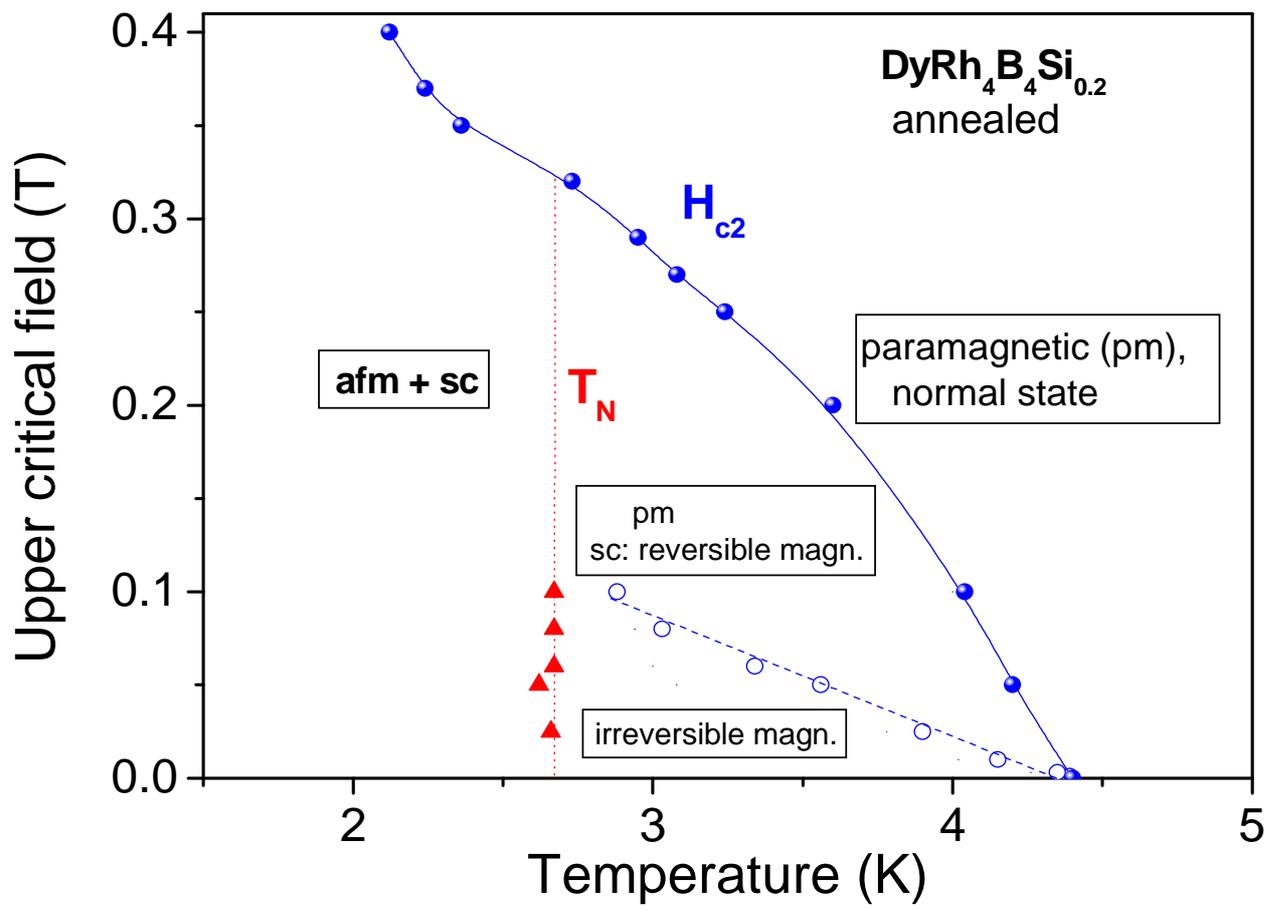

**Fig. 4**

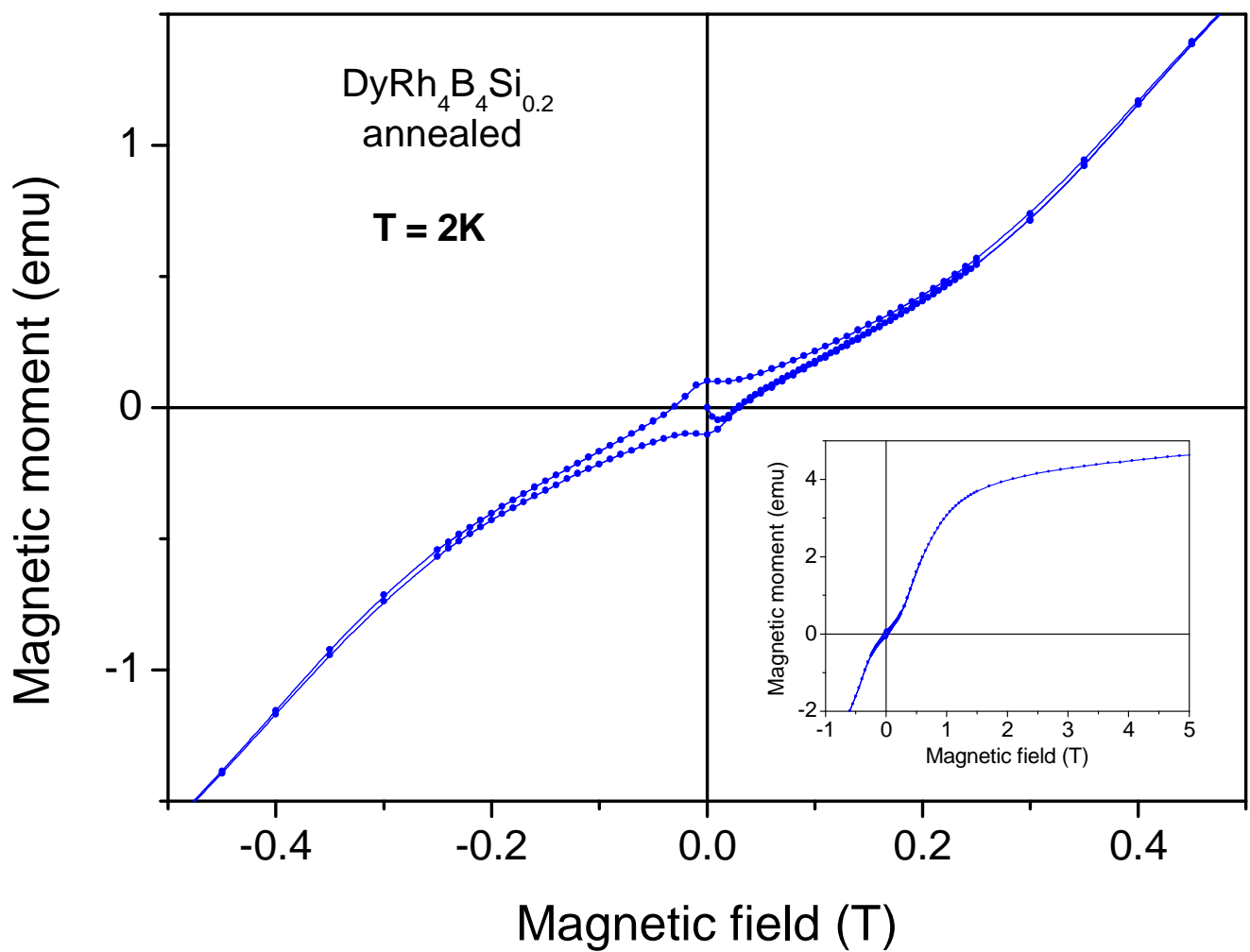

**Fig. 5**

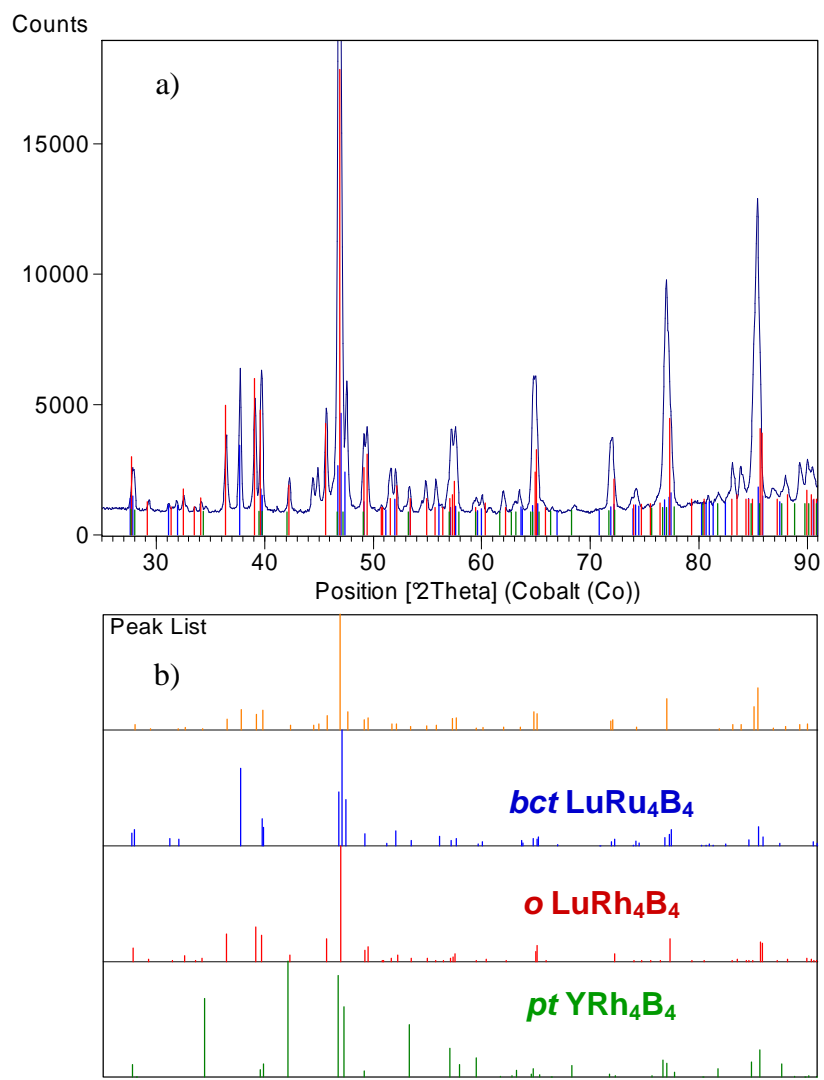

**Fig. 6**